\documentclass[twocolumn,showpacs,preprintnumbers,amsmath,amssymb]{revtex4}
\usepackage{graphicx}
\usepackage{dcolumn}
\usepackage{bm}

\begin{document}

\title{ From geodesic flow on a surface of negative curvature\\ to electronic
generator of robust chaos}

\author{Sergey P. Kuznetsov }
\altaffiliation{Kotel'nikov's Institute of Radio-Engineering and Electronics of RAS, Saratov
Branch, \newline
Zelenaya 38, Saratov 410019, Russian Federation}

\date{\today}

\begin{abstract}
Departing from the geodesic flow on a surface of negative curvature as a
classic example of the hyperbolic chaotic dynamics, we propose an electronic
circuit operating as a generator of rough chaos. Circuit simulation
in NI Multisim software package
and numerical integration of the model equations are provided. Results of
computations (phase trajectories, time dependences of variables, Lyapunov
exponents and Fourier spectra) show good correspondence between the chaotic
dynamics on the attractor of the proposed system and of the Anosov dynamics
for the original geodesic flow.
\end{abstract}

\maketitle

The hyperbolic theory is a part of the theory of dynamical systems
delivering a rigorous justification of the possibility of chaotic behavior
of deterministic systems both for the discrete-time case (iterative maps --
diffeomorphisms) and for the continuous time case (flows)~\cite{1,2,3,4,5}. The objects
of study are uniformly hyperbolic invariant sets in the phase space composed
exclusively of saddle trajectories. For conservative systems the hyperbolic
chaos is represented by the Anosov dynamics when the uniformly hyperbolic
invariant set either occupies a compact phase space (for diffeomorphisms),
or occupies completely a surface of constant energy (for flows). For
dissipative systems the hyperbolic theory introduces a special kind of
attracting invariant sets, the uniformly hyperbolic chaotic attractors.

A fundamental mathematical fact is that the uniformly hyperbolic invariant
sets possess the property of roughness, or structural stability~\cite{6}. It
means that the nature of the dynamics is robust and persists under small
variations of the system. Such systems should be of preferable interest to
any practical applications of dynamical chaos due to insensitivity to variation
of parameters, manufacturing imperfections, interferences, etc.~\cite{7,8,9}.
However, consideration of numerous examples of chaotic systems occurring in
different fields in nature does not justify the expectations regarding
occurrence of the hyperbolic chaos. In this situation, instead of looking
for "ready-for-use" examples it makes sense to turn to the purposeful
constructing the systems with hyperbolic dynamics appealing to tools of
physics and electronics~\cite{10,11} exploiting naturally the roughness
(structural stability). Namely, taking a formal example of hyperbolic
dynamics as the prototype, one can try to modify it in such way that the
dynamical equations become appropriate to be associated with a physical
system, hoping that due to the roughness the hyperbolic nature of the
dynamics will survive this transformation. In this article, departing from
the classical problem of the geodesic flow on a surface of negative
curvature, we propose an electronic device that operates as a generator of
robust chaos.

It is known that the free mechanical motion of a particle on a curved
surface is carried out along the geodesic lines of the metric, which is
defined by the quadratic form, expressing the kinetic energy $W$ via the
generalized velocities with coefficients depending on coordinates~\cite{12,13}.
In the case of negative curvature, the motion is characterized by
instability with respect to transverse perturbations. Therefore, if it
occurs in a compact domain, it appears to be chaotic~\cite{13}.

As an example,
consider the geodesic flow on the so-called Schwartz primitive surface~\cite{14},
which is defined in the three-dimensional space $(\theta _1 ,\,\,\theta _2
,\,\,\theta _3 )$ by the equation

\begin{equation}
\label{eq1}
\cos \theta _1 + \cos \theta _2 + \cos \theta _2 = 0,
\end{equation}

\noindent
and the motion takes place with constant kinetic energy

\begin{equation}
\label{eq2}
W = \textstyle{1 \over 2}(\dot {\theta }_1^2 + \dot {\theta }_2^2 + \dot
{\theta }_3^2 ).
\end{equation}

Here the mass is taken as a unit, and the relation (\ref{eq1}) may be regarded as
the imposed holonomic mechanical constraint. Because of the periodicity in
three axes, the variables $\theta _{1,2,3}$ may be defined modulo 2$\pi $,
and we can interpret the motion as proceeding in a compact domain, the cubic
cell of size 2$\pi $.

For curvature in this case it is possible to obtain an explicit
expression~\cite{15,16,17}.

\begin{equation}
\label{eq3}
K = - \frac{1}{2}\frac{\cos ^2\theta _1 + \cos ^2\theta _2 + \cos ^2\theta
_3 }{\left( {\sin ^2\theta _1 + \sin ^2\theta _2 + \sin ^2\theta _3 }
\right)^2}.
\end{equation}

With exception of eight points, where the numerator is zero, the curvature
$K$ is everywhere negative, so that the geodesic flow implements the Anosov
dynamics.

The dynamics associated with the geodesic flow on the surface (\ref{eq1})
occurs, for example, in the triple linkage mechanism of Thurston -- Weeks --
MacKay -- Hunt ~\cite{18, 15} in some special asymptotic case ~\cite{15,16,17}. It is also
of interest in the context of model description of a particle motion in
three-dimensional periodic potential, say, in the solid-state physics ~\cite{15,19}.

Using the standard procedure for mechanical systems with holonomic
constraints ~\cite{20}, we can write down the equations of motion in the form

\begin{equation}
\label{eq4}
\ddot {\theta }_1 = - \Lambda \sin \theta _1 ,\,\,\ddot {\theta }_2 = -
\Lambda \sin \theta _2 ,\,\,\ddot {\theta }_3 = - \Lambda \sin \theta _3 ,
\end{equation}

\noindent
where the Lagrange multiplier $\Lambda $ has to be determined with taking
into account the algebraic condition of mechanical constraint complementing
the differential equations. In our case

\begin{equation}
\label{eq5}
\Lambda = - {{\dot {\theta }_1^2 \cos \theta _1 + \dot {\theta }_2^2 \cos
\theta _2 + \dot {\theta }_3^2 \cos \theta _3}\over{\sin ^2\theta _1 + \sin
^2\theta _2 + \sin ^2\theta _3}}.
\end{equation}

Figure 1a shows a typical trajectory in the configuration space, which
travels on the two-dimensional surface (\ref{eq1}). The opposite faces of the cubic
cell are naturally identified, resulting is a compact manifold of genus 3;
in other words the surface is topologically equivalent to the "pretzel with
three holes" ~\cite{18,15}. Visually, you can conclude about chaotic nature of
the trajectory covering the surface in ergodic manner. The power spectrum of
the signal generated by the motion of the system is continuous, which is an
intrinsic feature of chaos (Fig.1b).

Taking into account the imposed mechanical constraint, there are four
Lyapunov exponent characterizing the behavior of perturbations about the
reference phase trajectory: one positive, one negative and two zero. One
exponent equal to zero appears due to the autonomous nature of the system;
it corresponds to the perturbation vector tangent to the phase trajectory.
Another one is associated with a disturbance of energy. Since the system
does not possess any certain characteristic time scale, the Lyapunov
exponents responsible for the exponential growth or decay of perturbations
are proportional to the velocity, i.e. $\lambda = \pm \kappa \sqrt W $,
where the coefficient is determined by the average curvature of the metric.
Empirically, from computations for the system under consideration $\kappa =
0.70$ ~\cite{16,17}.

\begin{figure}[htbp]
\centerline{\includegraphics[width=2.5in]{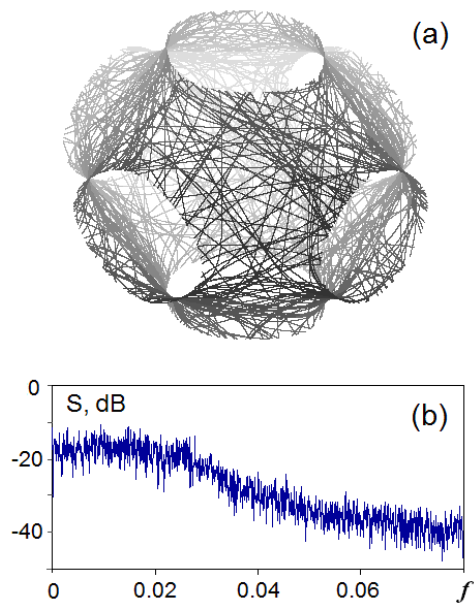}}
\label{fig1}
\caption{
Typical trajectory of the system (\ref{eq4}), (\ref{eq5}) in a three-dimensional
configuration space ($\theta _{1}$, $\theta _{2}$, $\theta _{3})$ (a)
and power spectrum of the variable $\dot {\theta }_1 $ for the motion with
the kinetic energy $W$ = 0.03 (b). Plotting the diagram (a) the angular
variables are related to the interval from 0 to 2$\pi $, i.e. it corresponds
to the fundamental cell, which is repeated with period 2$\pi $ along each of
the three coordinate axes.}
\end{figure}

In ~\cite{17} a self-oscillating system was suggested, where the sustained
dynamics corresponds approximately to the geodesic flow on the Schwarz
surface; there the kinetic energy is not constant but undergoes some
irregular fluctuations around a certain average level in the course
of the dynamics in time. This system is based
on three self-rotators, the elements whose state is defined by the angular
variables $\theta _{1,2,3}$ and generalized velocities $\dot {\theta
}_{1,2,3} $, and the steady motion of one element in isolation corresponds
to the rotation in either direction with a certain constant angular
velocity. The rotators are supposed to interact via the potential that is
minimal under the condition (\ref{eq1}). According to ~\cite{17}, in a certain range of
parameters the dynamics are hyperbolic, although for the modified system one
should speak about self-oscillatory chaotic regimes corresponding to hyperbolic
attractors rather then the Anosov dynamics. The purpose of this article is
to propose an electronic circuit implementation of such system and to
demonstrate its functioning as a generator of robust chaos.

For the construction of the electronic device the elements are required
similar to rotators in mechanics. Namely, the state of the element has to be
characterized by a generalized coordinate defined modulo 2$\pi $. An
appropriate variable of such kind is a phase shift in the voltage controlled
oscillator relative to a reference signal, like it is practiced in the
phase-locked loops ~\cite{21}.

Let us turn to the circuit diagram shown in Figure~2. The voltages
$U_{1,2,3}$ are used to control the phases of the oscillators V1, V2, V3, so
that the voltage outputs vary in time as $\sin (\omega t + \theta _{1,2,3}
)$, where the phases satisfy the equations $\dot {\theta }_i = kU_i ,\,\,\,i
= 1,2,3$, and $k$ is the coefficient characterizing the frequency control.
The center frequency of the oscillators is determined by the bias provided
by DC voltage source V4. The reference signal is generated by the AC voltage
source V5.

Assuming the output voltages of the multipliers A1, A2, A3 to be
$W_{1,2,3}$, for currents through the capacitors C1, C2, C3 we have $C\dot
{U}_i + (R^{ - 1} - g)U_i + \alpha U_i + \beta U_i^3 = R^{ - 1}W_i $, where
$i$=1,2,3, $C$=C1=C2=C3, $R$=R10=R11=R12, and $I(U) = \alpha U + \beta U^3$ is the
current-voltage characteristic of the nonlinear element composed of a pair
of the diodes. The equations take into account the negative conductivity $g
= R_2 / R_1 R_3 = R_5 / R_4 R_6 = R_8 / R_7 R_9 $ introduced by the elements
on the operation amplifiers U1, U2, U3. The voltages $W_{1,2,3}$ are obtained
by multiplying the signals
$\sin (\omega t + \theta _{1,2,3} )\cos \omega t$
from outputs of A4, A5, A6 by an output signal $W$ of the inverting
summing-integrating element containing the operational amplifier U4.

Input signals for the summing-integrating element are the output voltages of
the multipliers A7, A8, A9, so, that with account of the leakage current
through the resistor R16, we have $C_0 \dot {W} + r^{ - 1}W = - R_0^{ - 1}
[U_1 \sin (\omega t + \theta _1 ) + U_2 \sin (\omega t + \theta _2 ) + U_3
\sin (\omega t + \theta _3 )]\cos \omega t$, where C$_{0}$=C4,
$R_{0}$=R13=R14=R15, $r$=R16.

Using variables $\tau = t / \sqrt {4RCR_0 C_0 } ,\,\,\,u_i = 2k\sqrt {RCR_0
C_0 } U_i ,\,\,\,w = 2kR_0 C_0 W$ and parameters $\Omega = \sqrt {4RCR_0 C_0
} \omega ,\,\,\,\mu = (gR - \alpha R - 1)\sqrt {4R_0 C_0 / RC} ,\,\,\nu =
\beta / \sqrt {4k^4RC^3R_0 C_0 } ,\,\,\,\gamma = \sqrt {4RCR_0 / r^2C_0 } $,
we rewrite the equations in dimensionless form, where the dot means now the
derivative over $\tau $:

\begin{equation}
\label{eq6}
\begin{array}{l}
 \dot {\theta }_i = u_i ,\,\,i = 1,\,2,\,3, \\
 \dot {u}_i = \mu u_i - \nu u_i^3 + 2w\sin (\Omega \tau + \theta _i )\cos
\Omega \tau ,\, \\
 \,\dot {w} = - \gamma w - 2\sum\nolimits_{i = 1}^3 {u_i \sin (\Omega \tau +
\theta _i )\cos \Omega \tau } . \\
 \end{array}
\end{equation}

Non-trivial self-oscillatory behavior takes place at $\mu >0$; this
parameter may be varied by simultaneous tuning the resistances R1, R4,
R7.

Taking into account that $\Omega \gg 1$ one can simplify the equations
assuming that $u_{i}$ and $w$ vary slowly on the high-frequency period. Namely,
we perform averaging in the right-hand parts setting
\begin{equation}
\label{eq6a}
\overline {\sin
(\Omega \tau + \theta _i )\cos \Omega \tau } = \textstyle{1 \over 2}\sin
\theta _i
\end{equation}
and arrive at the equations
\begin{equation}
\label{eq7}
\begin{array}{l}
 \dot {\theta }_i = u_i ,\,\,\,\dot {u}_i = \mu u_i - \nu u_i^3 + w\sin
\theta _i ,\,\,\,\,i = 1,\,\,2,\,\,3, \\
 \dot {w} = - \gamma w - (u_1 \sin \theta _1 + u_2 \sin \theta _2 + u_3 \sin
\theta _3 ). \\
 \end{array}
\end{equation}

Finally, supposing $\gamma \ll 1$ we can neglect the respective term in the
last equation and to integrate it with substitution of $u_{1,2,3}$ from the
first equation; then we obtain $w \approx \cos \theta _1 + \cos \theta _2 +
\cos \theta _3 $, and the final result corresponds exactly to the equations
of Ref. ~\cite{17}:

\begin{equation}
\label{eq8}
\ddot {\theta }_i = \mu \dot {\theta }_i - \nu \dot {\theta }_i^3 - (\cos
\theta _1 + \cos \theta _2 + \cos \theta _3 )\sin \theta _i ,\,\,i = 1,2,3.
\end{equation}

Figure 3 shows a sample of the signal $U_{1}$ copied from the virtual
oscilloscope screen when simulating the dynamics of the circuit in the NI
Multisim software package,
and the spectrum obtained with the virtual spectrum analyzer. Visually, the
signal looks chaotic, without any apparent repetition of forms. Continuous
spectrum indicates chaotic nature of the process.
It is characterized by slow
decrease of the spectral density with frequency and is of rather good
quality in the sense of lack of pronounced peaks and dips.

\begin{figure*}[htbp]
\centerline{\includegraphics[width=7in]{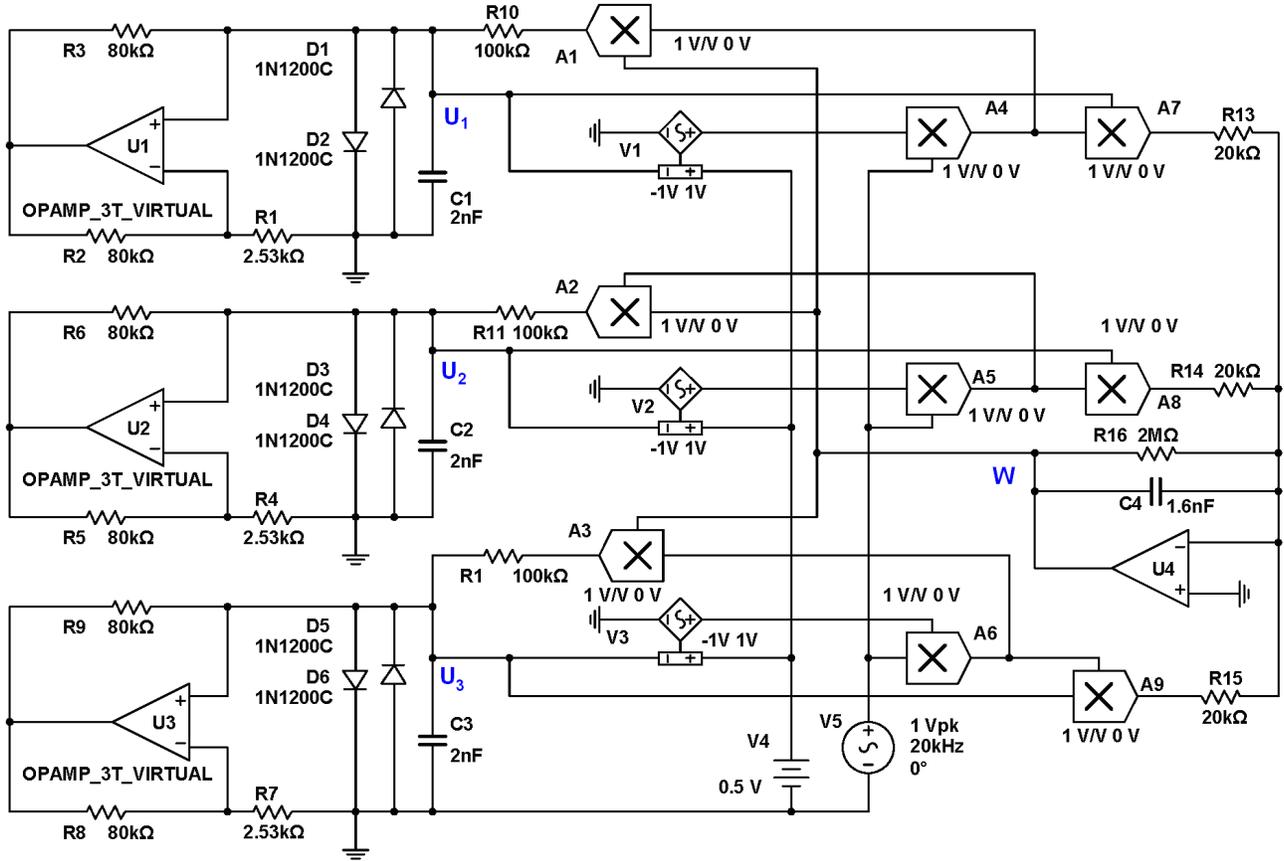}}
\label{fig2}
\caption{ Circuit diagram of the chaos generator in the Multisim software
package. The current-voltage characteristic of the nonlinear element with
two parallel contrarily directed 1N1200C diodes is given approximately by
the relation $I \approx \alpha U+\beta U^{3}$= 0.0039$U$+0.035$U^{3}$, where
the current is expressed in amperes and the voltage -- in volts. Coefficient
of frequency control for V1, V2, V3 is $k$/2$\pi $=40 kHz/V.}
\end{figure*}

\begin{figure*}[htbp]
\centerline{\includegraphics[width=7in]{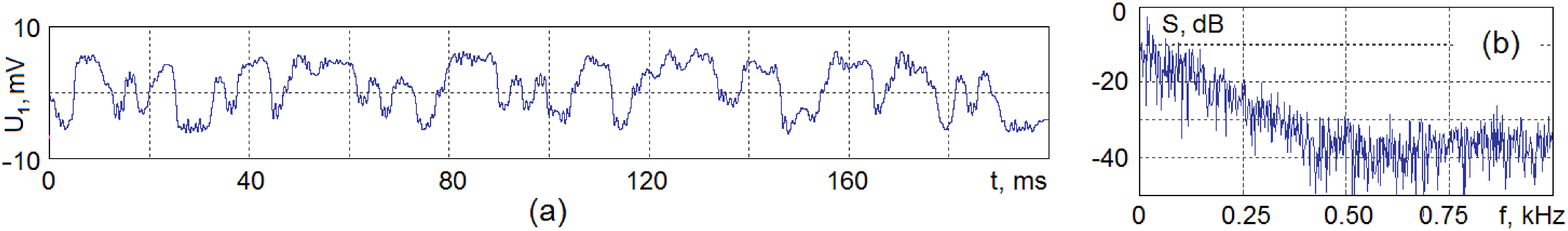}}
\label{fig3}
\caption{Voltage on capacitor C1 versus time in a sustained regime (a) and
its power spectrum (b) as obtained by circuit simulation in Multisim.}
\end{figure*}

\begin{figure*}[htbp]
\centerline{\includegraphics[width=7in]{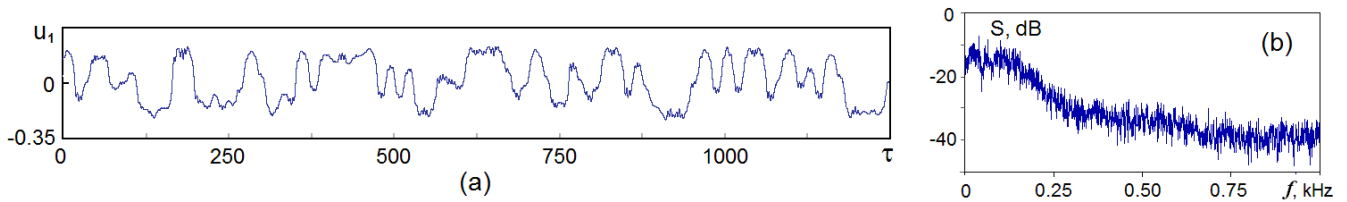}}
\label{fig4}
\caption{ Time dependence (a) and spectrum (b) of the variable $u_{1}$
obtained from numerical integration of Eqs.~(\ref{eq6}).}
\end{figure*}

In a frame of the circuit simulation it is difficult to explore some
characteristics, such as Lyapunov exponents, therefore, we turn to
comparison of the results with the model (\ref{eq6}), for which the relevant
analysis in the computations can be performed. Using the component nominals
indicated in the circuit diagram of Fig.2 and applying the conversion
formulas to the dimensionless quantities, we evaluate the parameters in the
equations (\ref{eq6}): $\mu = 0.07497,\,\,\nu = 1.73156,\,\,\gamma =
0.05,\,\,\,\Omega = 20.1062$. Figure 4 shows a plot of the dimensionless
variable $u_{1}$ versus time obtained from the numerical integration of the
equations (\ref{eq6}) (a), and the Fourier spectrum (b). The scales on the axes are
chosen to provide correspondence with Fig.3. Similar in form and
characteristic scales are samples of time dependences and spectra obtained
for the models (\ref{eq7}) and (\ref{eq8}).

\begin{figure}[htbp]
\centerline{\includegraphics[width=2.2in]{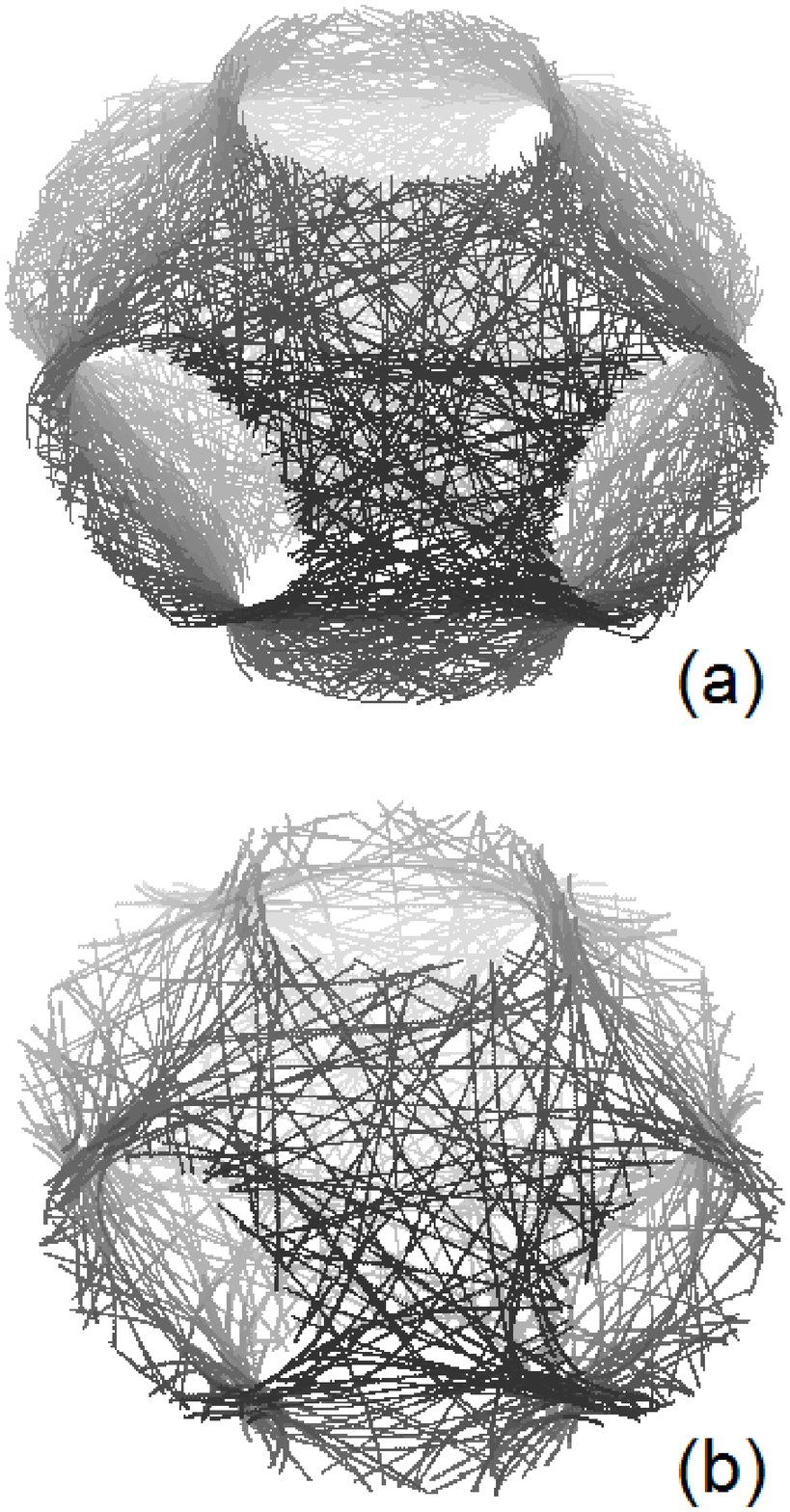}}
\label{fig5}
\caption{ Trajectories in the three-dimensional space $(\theta _1 ,\,\theta
_2 ,\,\theta _3 )$ for the model system (\ref{eq6}) (a) and for the electronic
device obtained from results of simulation in Multisim according to the
method described in the text (b).}
\end{figure}

As one can see, the dynamics of the electronic device are similar to the
original geodesic flow on the surface of negative curvature in the sense
that the trajectories in the space of coordinate variables ($\theta
_{1}$,$\theta _{2}$,$\theta _{3})$ are close to the Schwartz surface.
This is illustrated in Fig.5, which shows a trajectory found by numerical
integration of the equations for the model (\ref{eq6}), and a diagram obtained from
data of circuit simulation in Multisim. To plot the last one, the circuit
was complemented by three special signal processing modules. The output
signal of each of the voltage controlled oscillators subjected to
multiplication by $\sin \omega t$ and $\cos \omega t$, and after filtration
and separation of the low frequency components three pairs of the resulting
signals ($x_{k}$, $y_{k})$, $k$=1,2,3 were recorded in a file for subsequent
processing. According to the recorded data, at each time point three
variables defined modulo 2$\pi $ are evaluated as $\theta _k = \arg (x_k +
iy_k )$, $k$=1,2,3, and respective points are plotted. These diagrams can be
compared with Fig.1 for the geodesic flow on the surface of negative
curvature. Figure 5 shows that the trajectory remains close to the Schwartz
surface, though it is not located exactly on it; the pictures are "fluffed"
in the transverse direction. This effect becomes more pronounced with
increasing parameter $\mu $, as we move away from the critical point of
appearance of chaotic self-oscillations at $\mu $ = 0.

Figure 6 shows plots for all seven Lyapunov exponents calculated using
Benettin algorithm ~\cite{22,23,10} for the model (\ref{eq6}) depending on the parameter
$\mu $. In the presented range of $\mu $ we have one positive exponent,
other two are close to zero, and the rest are negative. The dependence on
the parameter is smooth, without pronounced peaks and dips, indicating the
roughness of the chaotic attractor. Note that in ~\cite{17} special calculations
were carried out based on verification of the absence of tangencies between
stable and unstable subspaces of perturbation vectors nearby a typical
trajectory on the attractor for the model (\ref{eq8}); it argues in favor of
assumption of the hyperbolic nature of the dynamics for the system under
consideration.

Particularly, at $\mu = 0.07497$ the Lyapunov exponents of the attractor are
$\lambda _1 = 0.1421\pm 0.0012$, $\lambda _2 = 0.0005\pm 0.0003$, $\lambda
_3 = 0.0000\pm 0.0002$, $\lambda _4 = - 0.0547\pm 0.0006$, $\lambda _5 = -
0.0582\pm 0.0009$, $\lambda _6 = - 0.1382\pm 0.0004$, $\lambda _7 = -
0.1591\pm 0.0022$, where errors indicated are the standard deviations
obtained under averaging data for 10$^{2}$ samples of duration $\tau $=5$
\cdot $10$^{4}$. The averaged dimensionless kinetic energy in this case
according to the computations is $W = \textstyle{1 \over 2}\overline {(u_1^2
+ u_2^2 + u_3^2 )} \approx 0.0425$, so, for the comparable geodesic flow the
Lyapunov exponents should be equal to $\pm 0.7\sqrt W \approx \pm 0.144$;
that agrees well with $\lambda _{1}$ and $\lambda _{7}$ relating to the
model (\ref{eq6}).

\begin{center}
\begin{figure}[htbp]
\centerline{\includegraphics[width=3.2in]{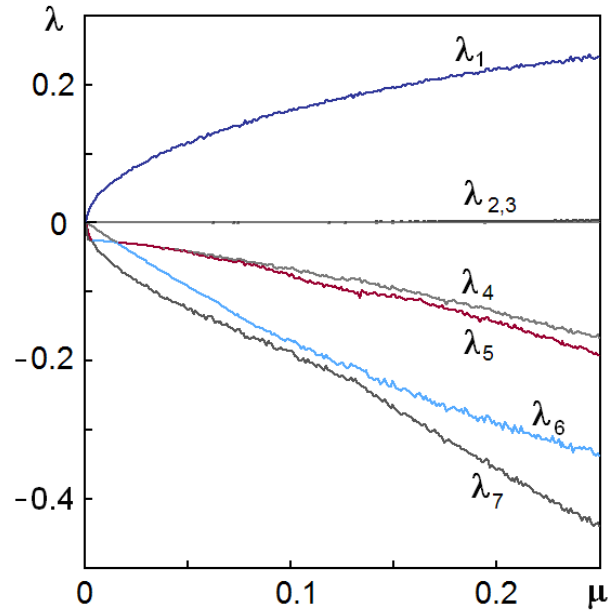}}
\label{fig6}
\caption{Lyapunov exponents of the system (\ref{eq6}) depending on the parameter of supercriticality $\mu $.}
\end{figure}
\end{center}

Concluding, in this paper a construction of the electronic generator of rough chaos is
proposed inspired by the problem of the geodesic flow on a surface of
negative curvature, which implements hyperbolic dynamics of Anosov. An
electron analog circuit simulation is provided in the NI Multisim software
package.
Also, the set of equations is derived to describe the system, and computational
study of chaotic dynamics is performed on the base of these equations. In
contrast to the previously considered electronic circuits with hyperbolic
attractors ~\cite{10,11,24,25}, in this case the hyperbolic dynamics is characterized
by higher degree of uniformity in expansion and compression for elements of
the phase volume in the course of evolution in continuous time. Thus, the
generated chaos has rather good quality of the power spectral
density distributions.

Although the particular circuit described in the article operates in the low
frequency range (kHz), it seems possible to implement similar devices
at high frequencies as well.

Since the hyperbolic dynamics are characterized by roughness, or structural
stability, as the mathematically proven attribute, it seems preferable for
practical applications of chaos due to low sensitivity to parameter
variations, various imperfections, noise, etc.

\begin{acknowledgments}
This work was supported by RFBR grant No 16-02-00135.
\end{acknowledgments}

\end{document}